\documentclass[aps,pra,reprint,amsmath,amssymb, floatfix,showkeys]{revtex4-2}

\usepackage[english]{babel}
\usepackage[utf8]{inputenc}
\usepackage[colorinlistoftodos, color=green!40, prependcaption]{todonotes}
\usepackage{graphicx}
\usepackage{bm}

\usepackage{amsmath}
\usepackage{xcolor}

\begin{document}


\title{{Mesoscopic Theory of Wavefront Shaping to Focus Waves inside Disordered Media} }


\author{Bart A. van Tiggelen} 
\affiliation{Universit\'e Grenoble Alpes, 
Centre National de la Recherche Scientifique (CNRS), 
Laboratoire de Physique et de Mod\'elisation des Milieux Condens\'es (LPMMC), 
38000 Grenoble, France} 
\email[]{Bart.Van-Tiggelen@lpmmc.cnrs.fr}

\author{Ad Lagendijk} 
\affiliation{Complex Photonic Systems (COPS), MESA+ Institute for Nanotechnology, University of Twente, P.O. Box 217, 7500 AE Enschede, The Netherlands} 

\author{Willem L. Vos}
\affiliation{Complex Photonic Systems (COPS), MESA+ Institute for Nanotechnology, University of Twente, P.O. Box 217, 7500 AE Enschede, The Netherlands} 

\affiliation{Universit\'e Grenoble Alpes, 
Centre National de la Recherche Scientifique (CNRS), 
Laboratoire de Physique et Mod\'elisation des Milieux Condens\'es (LPMMC), 
38000 Grenoble, France} 


\date{ 14 october 2024 } 

\begin{abstract}
We describe the {theory of focusing waves} to a predefined spatial point {inside} a disordered {three-dimensional medium} by the external shaping of {$N$} different field sources outside the medium, {also known as wavefront shaping}. 
We {derive} the energy density of the wave field {both} near the focal point and anywhere else inside the medium, {averaged over realizations\textit{ after} focusing}. 
{To this end, we conceive of a point source at the focal point that emits waves to a detector array that - by time reversal - emits the desired shaped fields. }
{It appears that the energy} density is formally equal to intensity speckle described by {the so-called} $C_1$, $C_2$, $C_3$ and even $C_0$ {correlations} in mesoscopic transport theory, {yet the density also obeys a diffusion equation}. 
The $C_1$ {correlations} describes the focusing in the random medium very well, but do not generate a new source of energy that {is conceived} at the focal point. 
A source emerges {only} when the $C_2$ speckle is incorporated. 
The role of $C_0$ speckle, describing fluctuations in the {local density of optical states (LDOS)} is also investigated, {but hardly plays a role in the focusing. } 
Finally, we use the {concept of an energy source inside the medium} to model the {well-known} optimized transmission by a slab using wavefront shaping. 
\end{abstract}


\maketitle

\section{Introduction}\label{sec:introduction}
Waves can be focused to predefined points by the manipulation of incident wave fronts~\cite{Vellekoop2007OptLt}. 
This new method is now well known in optics as ''wavefront shaping" (WFS)~\cite{Mosk2012NP}. 
In a disordered medium, WFS is efficient because multiple scattering is characterized by short range correlations that can suppress statistical fluctuations. 
In this procedure, with many potential applications, one manipulates the phases and amplitudes of spatial or spectral elements of an incident wave packet to optimize either the focusing to a predefined point, the total transmission, or even the delay time~\cite{Rotter2017RMP}. 
Recent work concentrated not just on the focusing but also on the energy density of the waves inside the medium  after wave front shaping. 
It was found that after optimizing transmission, the density profile inside the medium is dominated by the first eigenmode of the diffusion equation~\cite{Ojambati2016OE}, {as confirmed by Davy \textit{et al.}~\cite{Davy2015NatComm} and Koirala \textit{et al.}~\cite{Koirala2017PRB}. } 
This surprising feature is remarkable yet no simple explanation has been given to date. 

\begin{figure*}[htbp]
\includegraphics[width=0.9\textwidth,angle=0]{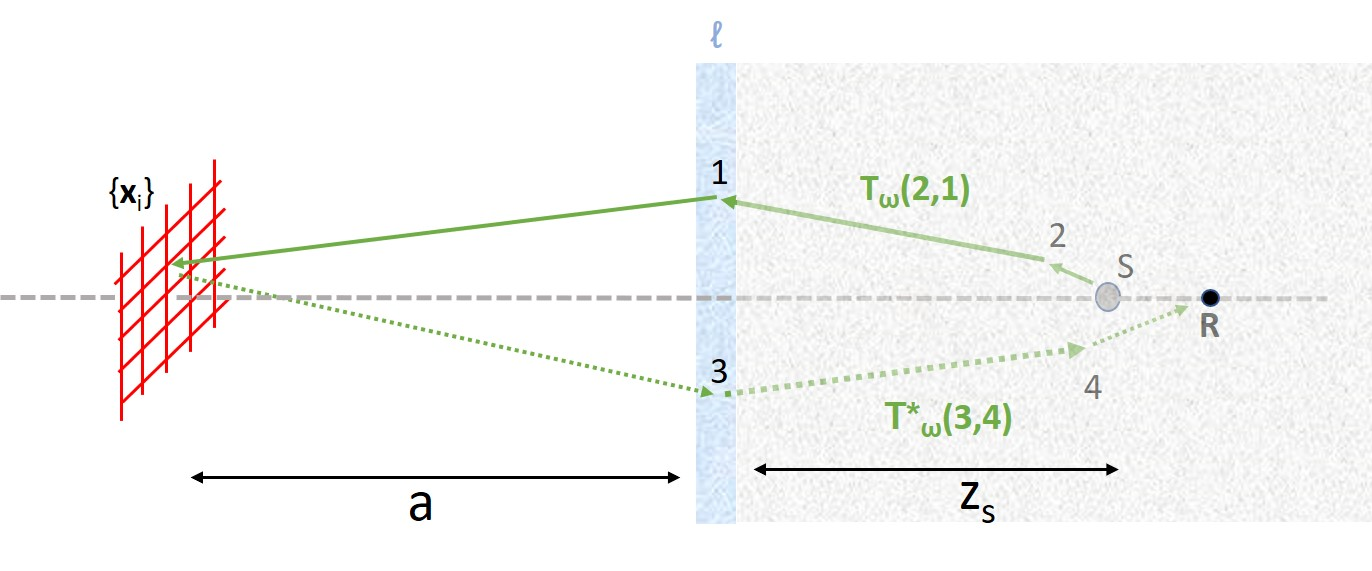}
\caption{Schematic of wavefront shaping to focus waves to a point $S$ inside a disordered 3D half-space on the right. 
The medium has a skin layer with a depth equal to about one mean free path $\ell$. 
A wave packet from a virtual point source at $S$ is sent towards the detector array with $N$ detectors at positions $\{ \mathbf{x}_i\}$ in the ''far field" of the medium, meaning here that $a \gg z_S$. 
The optimal shaping is established by time-reversal of the entire wave packet. 
It necessarily refocuses on a diffraction-limited spot $R \approx S$ with a background energy that is calculated in this work for all positions $R$. 
The path of the wave from source to array is separated into 3 parts: a) release of energy by a virtual source at focus $S$ to a nearby scatterer $2$, b) radiative transport towards a position $1$ in the skin layer with $T$-matrix $T_\omega(2,1)$, and c) ballistic propagation from skin layer towards array. The same is done for the time-reversed signal, here equal to the complex conjugate and shown as a dashed line. }
\label{fig:setup_WFS}
\end{figure*}
In this work we analyze the WFS to  focus an incident wave packet to a given point deep in a disordered slab or half-space. 
In Figure~\ref{fig:setup_WFS} we show the set-up for a disordered half-space. We use a mesocopic theory with a statistical approach to deal with wave fluctuations. This approach does not reveal information on individual realizations, such as the specific realization that optimizes some observable. Nevertheless, given a specific realization of the disorder, the wave front to be emitted by an array of sources designed to optimally focus to a point $S$ inside the random medium can be obtained by first sending a wave packet from a virtual source at $S$, and time-reverse the received signal at the array. 
Previous work has established that the focusing is stabilized by short-range correlations and close to the statistical average~\cite{vanTiggelen2003PRL}. The important role of non-Gaussian mesoscopic fluctuations has also been addressed~\cite{Hsu2017NatPhys}.

We will assume a monochromatic wave packet and an array with angular opening $\Omega_A$. 
This array sends plane waves with propagation directions inside the solid angle $\Omega_A$ when the phases emitted by the different elements $\{ \mathbf{x}_i\}$ vary linearly. When they are manipulated otherwise, many complicated wave fronts can be achieved. Here we want to employ WFS to focus to a point  $S$ in the medium. The time-reversal method implicitly generates an ideal candidate, not necessarily unique. In addition, scattering from disorder generates an background energy density of the wave field that varies with the position of point $R$ in Figure~\ref{fig:setup_WFS}. Intuitively,  we may expect the focal point $S$ to become a secondary source of energy for diffuse waves in the random medium and contribute to this background energy. 
We will analyze different correlations of type $C_0$ (fluctuations in density of states), $C_1$ (Gaussian speckle of intensity) and $C_2$ (correlations in total transmission) known in mesoscopic wave physics, and show that $C_2$ makes a dominant contribution to the energy density of the waves away from focal spot, and also generates this expected source.

Because diffusion is taken to be isotropic, the typical diffuse halo created by a virtual source at depth $z_S$ has size $z_S$. 
We will also assume that the distance $a$ of the array satisfies $a \gg z_S$. 
In that case, waves arriving at points $1$ and $3$ in Figure~\ref{fig:setup_WFS} still have approximate normal incidence. 
The average propagation from array to skin layer is described by the Green function 
\begin{eqnarray}\label{Gskin}
    G(i, 1) &=& - \frac{1}{4\pi |\mathbf{r}_i - \mathbf{r}_1|} e^{ ik |\mathbf{r}_i - \mathbf{r}_1|} e^{-z_1/ 2\ell \cos \theta} \nonumber \\ &\approx &
     - \frac{1}{4\pi a} e^{ika}e^{-ik \bm{\rho}_1 \cdot\mathbf{ x}_i/a} e^{-z_1/ 2\ell}
\end{eqnarray}
Near points $R$ and $S$ deep in the medium we adopt the bulk expression for average wave propagation,
\begin{eqnarray}\label{Gbulk}
    G(S, 2) &=& - \frac{1}{4\pi |\mathbf{r}_S - \mathbf{r}_2|} e^{ i (k+i/2\ell)  |\mathbf{r}_S - \mathbf{r}_2|}
\end{eqnarray}
which restricts the position of point $2$ within a mean free path $\ell$ from $S$.

\section{Time-Reversal to Optimize Wave Front}
Let us briefly recall the process of time-reversal, adapted to the present simplified context. 
For scalar waves, time-reversal reduces to complex conjugation. When the virtual source at $S$ starts releasing  its energy at $t=0$, the field is registered outside the medium on the array between times  $0 < t < T$, sent back time-reversed immediately and measured at point $R$ at time $\tau > T$. We assume that the signal has a  spectral bandwidth $B$ small enough to be consider quasi-monochromatic but that $T \gg 2\pi /B$. The field sent back into the medium and arriving at $R$ is then given by~\cite{Derode1995PRL,papa}

\begin{eqnarray}\label{TRM}
    \Phi(R, \tau+ 2 T) &=& \int_{-\infty}^{\infty} \frac{d\omega}{2\pi}  e^{-i\omega \tau} s(\omega)\nonumber \\
     &\times &  G_{S\rightarrow \{\mathbf{x}_i\}}(\omega+i0)G_{ \{\mathbf{x}_i\}\rightarrow R }^*(\omega+i0)
\end{eqnarray}
From previous work we know that this signal typically contains an arrival at $\tau <0$ and - by reciprocity - one at $\tau > 0$ associated with  waves traveling from $S$ to $R$ and vice-versa. In the frequency domain, we can split this propagation  up  into different parts as shown in Figure~\ref{fig:setup_WFS},

\begin{eqnarray}\label{TRS}
    \Phi(R, \omega ) = &D& \sum_{i=1}^N \int_{1,2,3,4}  G(S,2) T(2,1) G(1,i) \nonumber \\
    &\times & G^*(i,3) T^*(3,4) G^*(4,R)
\end{eqnarray}
with all factors taken at equal frequency, therefore the label is omitted. 
We have summarized the ''virtual" source intensity of $S$ and any possible loss or gain factor in time-reversal by a complex number $D$. Hereafter we shall put $F= D/(4\pi a)^2$ as the typical amplitude of the incident wave front arriving at the boundary (with same unit as the field itself). 
The quality of focusing is determined by 1) the size of the focal spot around $S$, and 2) the ratio of peak value in energy at $S$ and surrounding ``average background". 
As mentioned earlier, we expect this background to contain a component associated with a real source at $S$. 
In time-reversal experiments, a large enough bandwidth (larger than the Thouless frequency proprtional to $D/z_s^2$ that determines uncorrelated frequencies) guarantees self-averaging and stability of the process. In this work this is provided by the opening of the  array.

\subsection{$C_1$ correlations}
The ensemble-average of the energy density  $\langle |\Phi(R)|^2\rangle$ contains \emph{four} T-matrices. 
In the well-known $C_1$ approximation we have Gaussian decoupling of complex conjugates~\cite{vanRossum1999RMP},
\begin{eqnarray}\label{C1}
&&  \langle T(2,1) T^*(3,4) T^*(2', 1') T(3',4') \rangle =  \ \ \ \ \nonumber \\
& &   \ \ \ \   \langle T(2,1) T^*(3,4) \rangle \langle T^*(2', 1') T(3',4') \rangle  \nonumber \\
&&  \ \ \ \ +  \langle T(2,1) T^*(2',1') \rangle \langle T^*(3, 4) T(3',4') \rangle
\end{eqnarray}
 where the accents refer to the positions in the path followed by the time-reversed path. The first term is just equal to the average field  squared and peaks at $R \approx S$. This follows from the equality

\begin{equation} \label{eq:peak0}
    \int_2 G(S,2) G^*(2,R) = \frac{\ell}{4\pi } \frac{\mathrm{Im } \, G( S,R)  }{ \mathrm{Im }\, G( 0) } \equiv \frac{\ell}{4\pi } P(S,R)
\end{equation}
which describes a diffraction-limited peak near $S$ oscillating at the scale of the wavelength and decaying exponentially with the mean free path $\ell$. 
In the diffusion approximation we have
\begin{equation}\label{DA}
    \langle T(2,1) T^*(3,4) \rangle = L(1,2) \left[ \delta_{13}\delta_{24} +  \delta_{12}\delta_{34}  \right]
\end{equation}
with the propagator $L$ obeying the stationary diffusion equation 
\begin{equation}\label{DA2}
  - \frac{1}{3} \ell^*\bm{ \nabla}^2_1 L(1,2)  =  \frac{4\pi}{\ell^2 }\delta_{12}
 \end{equation}
where $\ell^*$ is the transport mean free path. To simplify, we will assume $\ell^* =\ell$ in the rest of this work. The two terms in Eq.~(\ref{DA}) are imposed by reciprocity but the second "Coherent Backscattering " term only contributes when point $1$ is close to point $2$, and is here excluded because $1$ is located in the skin layer and $2$ is close to $S$ deep in the bulk. 
With the approximations explained above we obtain for the peak energy density $P_1$ (averaging brackets are dropped for simplicity),
\begin{eqnarray*}\label{eq:peak_energy}
P_1(R,S)  &=&  |F|^2 N^2 \left(\frac{\ell}{4\pi } \right)^2 P( S,R)^2 \nonumber \\
    &\ \ &\times \left| \int_0^\infty  dz_1  e^{-z_1/\ell } L(z_S, z_1, \mathbf{q}=0) \right|^2
\end{eqnarray*}
with $N$ the number of elements in the array all contributing equally and coherently, and $\mathbf{q}$ the transverse Fourier component. 
Here, $\mathbf{q}=0$ appears because the effective area $\sim z_S^2$ illuminated by the virtual source is much smaller than the  area $\sim a^2$ available. 
If we impose the usual radiative boundary condition $L(-z_0, z) = 0 $ for all $z$, with $z_0\approx \frac{2}{3}\ell$, it follows that $L(z,z', \mathbf{q}=0) = (12\pi/\ell^3) [z_0 + \mathrm{min}(z,z') ]$. 
To avoid messy formulas especially later we will approximate the integrals $\int dz \exp(-z/\ell) f(z) \approx \ell f(0) $.
Then, for $z> 0$,
\begin{eqnarray}\label{peak}
 P_1(S,R)   =  9 |F|^2 N^2  P(S,R)^2  \left( \frac{z_0}{\ell}\right)^2
\end{eqnarray}
According to this expression, not valid for $S$ near the boundary, the focused energy spot is independent of the depth of $S$. 
We emphasize that the spot is not due to a local energy source (see Appendix~\ref{sec:Appendix_B_C2_peak}). 
In this steady state picture, the focus is due to constructive interference between incoming and outgoing waves at $S$ that causes the energy to accumulate locally, much like what happens in the focal spot after time-reversal~\cite{draeger}. 

The second term in Eq.~(\ref{C1}) is usually associated with short-range speckle correlations, but here  generates a "diffuse" background energy $B_1$, \textit{i.e.}, that depends on $R$ but without a distinguished peak near $S$. This background suffers from dephasing between the phase factors  $ \exp(i\bm{\rho}_i \cdot\bm { x}_i/a)$ in Eq.~(\ref{Gskin}),
\begin{eqnarray}\label{C1background}
&& B_1(z, {\bm \rho})   =  |F|^2 \frac{\ell^4}{(4\pi)^2 } \sum_{ij} L(0,z_S, \mathbf{q}_{ij}) L(0, z, \mathbf{q}_{ij}) e^{i\bm q_{ij}\cdot \bm \rho}\nonumber \\
 \end{eqnarray}
with the transverse wave vector $ \mathbf{q}_{ij} = k (\mathbf{x}_i -\mathbf{ x}_j)/a$ governing interference between different array elements. For finite $\mathbf{q}$, the diffusion propagator takes the form,
\begin{equation*}
    L(z,z', \mathbf{q}) = \frac{12\pi }{\ell^3}   \frac{ e^{-q|z-z'|} - e^{-q(z+z'+2z_0)}  }{2q}
\end{equation*}
If the typical distance between array elements obeys  $\Delta x < (\lambda/2\pi) a/z_s$, clearly true if $a\gg z_s$, the sum over $j$ reaches the continuum limit so that we can replace  by an integral, that in addition is independent on $i$ since for $z_S \gg \ell$ the propagator  $L(z_s,0,q)$ decays rapidly with $q$. We introduce $N (\Delta x)^2/a^2 = \Omega_A$ as the total angular opening of the array. 
We assume $\Omega_A < 1$ so that approximate normal incidence remains valid for all array elements without the need for $\Omega_A$ to be very small. For $z>0$ the expression for $ B_1  $ reduces to 
\begin{eqnarray}\label{C1background2}
    B_1(z, {\bm \rho } )   &=&  \frac{|F|^2N^2\ell^4}{(4\pi)^2 k^2\Omega_A}  \int d^2\mathbf{q} L(0,z_S,\bm q) L(0,z, \bm q) e^{i\bm q \cdot \bm \rho} \nonumber \\
     &=& 9|F|^2N^2 \frac{z_0^2}{\ell^2} \frac{2\pi z_0^2}{k^2\Omega_A }  \frac{z+z_S +2z_0}{[(z+z_S +2z_0)^2 + \rho^2]^{3/2}}\nonumber \\
\end{eqnarray}
For $S$ near the boundary this is proportional to the familiar  emission of a diffuse point source created by a focus from outside~\cite{johnC0}. For $S$ deep in the medium the energy density is roughly constant as long as $z, \rho < z_S$. 
The diffuse transverse halo has a typical  surface that  grows with depth according
\begin{equation}\label{surfaceC1}
    \langle\rho^2 (z,z_S)\rangle_1 \equiv \frac{\int d^2 {\bm \rho} \, B_1({z,\bm \rho})  }{B_1 ({z, \bm \rho}=0)} = 2 \pi (z+z_S+2z_0)^2
\end{equation}
and is small compared to $\pi a^2$ at the surface ($z=0$) when $z_S \ll a$, justifying an earlier assumption. 
We can compare the background energy density near $S=(z_S,{\bm \rho}=0)$ to the peak value given by Eq.~(\ref{peak}),
\begin{equation}\label{ratio}
     \frac{ B_1(R=S) }{   P_1(R=S)  } \approx \frac{2\pi/[2k(z_S+z_0)]^{2}}{\Omega_A}
\end{equation}
This expression shows that in the $C_1$ approximation the typical angular correlation for WFS to a point $S$ deep inside the half-space is $\delta \Omega  \sim 2\pi /(2kz_s)^2 $. 
The conclusion is that the finite angular coverage - even small -suppresses the $C_1$ background and makes the focus  prominent and stable against statistical fluctuation in the medium~\cite{vanTiggelen2003PRL}. 

In Appendix~\ref{sec:Appendix_A_current} we show that the background energy density $B_1$, despite being produced here by a preceding  time-reversal procedure, obeys the usual diffusion equation known for diffuse energy produced by normal incident radiation, in particular with the same diffusion constant $D$. 
The diffusion equation is also characterized by a source of energy, proportional to $-\nabla^2 B_1$. From Eq.~(\ref{C1background2}) we infer that this source is proportional to $L(0,z_s,\bm \rho) \delta(z) \sim \delta(z) \times z_S/(z_S^2 +\rho^2)^{3/2}$, \textit{i.e.}, only at the surface boundary and with a finite transverse width equal to $z_S$. Because no other source exists in the medium and also no flux is transmitted, the energy flux $F_1$ across any transverse surface given by,
\begin{equation}\label{current}
    F_1(z>0) = \int d^2 \bm\rho J(z,\bm\rho) = -D\partial_z  \int d^2 \bm\rho B_1(z,\bm\rho)=0,
\end{equation}
vanishes. 
Nevertheless, for any $z>0$ and for fixed $\bm\rho$, the current density itself does not vanish. 
Since the incoming flux vanishes due to the radiative boundary condition $B_1(-z_0)=0$, the totally reflected (average) flux is given by
\begin{eqnarray}\label{RC1}
    F_1(z<0) &= &-D \int d^2 \bm\rho \, \partial_z B_1(z<0, \bm\rho) \nonumber \\
        &= & -\frac{9|F|^2N^2}{(k\ell)^2} \frac{(2\pi)^2 z_0^2}{\Omega_A } \times \frac{c_0}{2}
\end{eqnarray}
with $z_0 = 2\ell/3$ and $D = c_0 \ell/3$. 
We need this outgoing flux later when we include $C_2$ to normalize the reflection coefficient.

Finally, one may wonder how nearby scatterers affect the focusing profile. 
They are not taken into account by the present approach. 
In Appendix~\ref{sec:Appendix_C_C0_correlations} we discuss these so-called $C_0$ correlations. 
The conclusion is that they perturb the peak profile and have relative weight $1/k\ell$ compared to the peak value (\ref{peak}). They do not contribute to background.

\section{$C_2$ correlation and source}\label{sec:C2_correlation}

\begin{figure*}[htbp]
\includegraphics[width=0.9\textwidth]{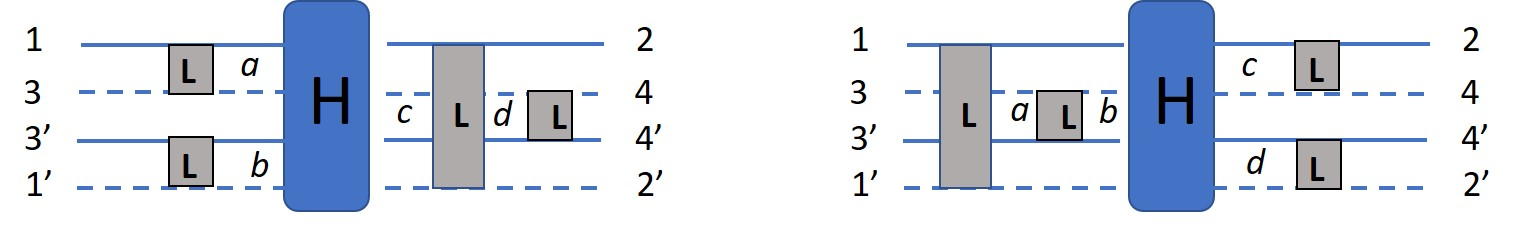}
\caption{Hikami diagrams for $C_2$ correlations in the average $\langle T(2,1) T^*(3,4) T^*(2',1')T(3',4') \rangle$ to focalization. Dashed lines denote complex conjugates.
The Hikami vertex $H$ exchanges momenta at position $\mathbf{H}$ in the medium, over which is to be integrated.}
\label{figm4}
\end{figure*}
In this section, we investigate the importance of $C_2$ correlations for the focusing of energy at point $S$ and its energy density around. 
The two diagrams are shown in Figure~\ref{figm4}. 
The figure on the left is expected to contribute to the background because its is proportional to $\delta_{13}\delta_{1'3'}\delta_{22'}\delta_{44'}$. 
This links both endpoints $4$ and $4'$ to the point of observation $R$ and does not decorrelate when $R$ changes (recall Figure~\ref{fig:setup_WFS}). In addition, since $1$, $3$ and $1'$, $3'$ are linked the diagram is also insensitive to phase differences between received and emitted waves by the array, as was the case for the peak energy in Eq.~(\ref{peak}). The figure on the right on the other hand is proportional to $\delta_{24}\delta_{2'4'}$ and only survives when $R$ and $S$ are separated by at most one mean free path and thus contributes to focalization. 
The decorrelation between different array elements makes it a factor $\Omega_A (kz_S)^2$ smaller.
The diagram on the left is expressed mathematically as~\cite{vanRossum1999RMP}. 

\begin{eqnarray}\label{C2}
&&  \langle T(2,1) T^*(3,4) T^*(2',1')T(3',4') \rangle_{C_2} =  \nonumber  \\
&& N_H \int d^3\mathbf{H} \left[ \bm{\nabla}_a \cdot \bm{\nabla}_b + \bm{\nabla}_c \cdot \bm{\nabla}_d  - \frac{1}{2} \sum_i \bm{\nabla}^2_i\right]_{abcd=\mathbf{H}} \nonumber \\
 &\ &  \ \   L(1,a)L(1',b)L(c,2) L(d,4) \, \delta_{22'}\delta_{44'}\delta_{13}\delta_{1'3'}
\end{eqnarray}
with $N_H = \ell^5/48\pi k^2$ and $\bm{H}$ the location of the Hikami vertex in the medium. The diffuse kernel $L$ was defined earlier in Eq.~(\ref{DA2}). With this, the background energy density at position $R$ is given by,
\begin{eqnarray*}
 && B_2(R) = |F|^2 N_H\int_{11'24} \int  d^3\mathbf{H}  \sum_{ij}^N \\
   && |G(2,S)|^2 |G(4,R)|^2 |G(1,\mathbf{x}_i)|^2 |G(1',\mathbf{x}_j)|^2\\
   & \times & \left[ \ {\bf\nabla}_\mathbf{H } L(1,\mathbf{H}) \cdot \bm{\nabla}_\mathbf{H } L(1',\mathbf{H}) L(\mathbf{H},2) L(\mathbf{H},4) \right.  \\
   &+& \left.  L(1,\mathbf{H}) L(1',\mathbf{H}) \bm{\nabla}_\mathbf{H} L(\mathbf{H},2) \cdot \bm{\nabla}_\mathbf{H} L(\mathbf{H},4) +\right. \\
 &&  \left. + \frac{1}{2} \sum_{i=abcd} -\bm{\nabla}_i^2 L_(1,a) L(1',b)  L(c,2) L(d,4) \right]_{abcd=\mathbf{H}}
\end{eqnarray*}
All phase factors in Eq.~(\ref{Gskin}) cancel so that all $N$ angular channels are correlated. 
We approximate $|G(2,S)|^2 \approx (\ell/4\pi) \delta_{2,S}$ which may have  to be reconsidered for $R \approx S$ in the last line in the expression above that will actually dominate.  
After some straightforward algebra we find the lengthy expression for the half-space

\begin{eqnarray}\label{eq:C2simple}
 && B_2(R)= |F|^2 N^2 N_H \frac{\ell^2}{(4\pi)^2} \times   \nonumber \\
 && \int_0^\infty  dz_1 e^{-z_1/\ell} \int_0^\infty  dz_{1'} e^{-z_{1'}/\ell  }  \nonumber   \\
   &\times &  \large\left[ - \frac{1}{2} \partial_{z_H} A(z_H=0,R,S) \tilde{L}(1,0 ) \tilde{ L}(1',0) \right. \nonumber \\
   &+& \frac{1}{2} \partial_{z_H} \left( \tilde{L}(1,z_H=0 ) \tilde{L}(1',z_H=0 ) \right) A(z_H=0,R,S) \nonumber \\
   &+& 2 \int_0^\infty dz_H  \partial_{z_H}\tilde{L}(1,z_H ) \partial_{z_H} \tilde{L}(1',z_H )  A(z_H,R,S) \nonumber \\
   &+& \left. \int d^3 \mathbf{H} \left( -\bm{\nabla}_c^2 -\bm{\nabla}_d^2 \right) \tilde{L}(1,z_H )\tilde{ L}(1',z_H )  L(c,S) L(d,R)\right] \nonumber \\
\end{eqnarray}
where we have written $\tilde{L}(z,z') = L(z,z', \mathbf{q}=0)$, defined $z_H$ as the $z$-component of the location $\mathbf{H}$ of the Hikami box, and introduced, for $R= (z, \bm\rho)$
\begin{eqnarray*}
  A(z_H,R,S) &=& \int d^2 {\bm \rho}_{H} L(z_H,z_S, {\bm\rho}_{ H} )  L(z_H,z, \bm\rho_{H} - \bm \rho  )\\
  &=& \int \frac{d^2\mathbf{q}}{(2\pi)^2} L(z_H,z_S, \mathbf{q} )  L(z_H,z, \mathbf{q}) e^{-i\bm q \cdot \bm \rho }
\end{eqnarray*}
which for $z_H \approx 0$ is proportional to the $C_1$ background in Eq.~(\ref{C1background2}). We will simplify the above expression dramatically. As done before we replace  the integrals over the skin layer by $\ell f(0)$ when possible.
In the first two terms of Eq.~(\ref{eq:C2simple}) the Hikami vertex is located exactly at the boundary. Their sum  is proportional to $ A(z_H =0,R,S) - 2z_0 \partial_{z_H} A(z_H=0,R,S) \approx 0$ since $A \sim (z_H + z_0)^2 $ for $z_H < 0$. In the third term  the Hikami box is situated in the skin-layer $0 < z_H < \ell$ since  the diffuse propagator in a half-space  $\partial_{z_H}\tilde{L}(1,z_H )$ vanishes for $z_H > z_1$. Therefore it is proportional to $A(z_H=0, R,S)$ and constitutes as such a correction to the  $C_1$ expression derived in Eq.~(\ref{C1background2}),

\begin{eqnarray}
B_2^{(3)}(R)= \frac{9 |F|^2 N^2 }{(k\ell)^2}\times 3 \pi A(0, R, S) = \frac{3 \Omega_A}{4\pi} B_1(R)
\end{eqnarray}
This confirms the angular coverage $\Omega_A/2\pi$ to be the leading parameter in our expansion of wave correlations.
Finally, in the last term the Hikami box is located either \emph{exactly} either at the virtual source $S$ or at the detection point $R$. 
The two diffusion operators generate $\delta_{\mathbf{H},\mathbf{R}} L(\mathbf{H},\mathbf{S}) + \delta_{\mathbf{H},\mathbf{S}} L(\mathbf{H},\mathbf{R})$. 
This contribution $B^{(4)}_2$ can be interpreted as a diffuse propagation from the slab boundary first to  $R$ and subsequently to $S$, or \textit{vice versa}. 
\begin{eqnarray}\label{C2final}
&& B_2^{(4)}(R) = \frac{9|F|^2 N^2}{(k\ell)^2} \frac{3z_0^2}{2\ell}  \times \nonumber \\
 \ \ \ &&  \left[\frac{1}{\sqrt{(z-z_S)^2 +\rho^2}} - \frac{1}{\sqrt{(z+z_S + 2z_0)^2 +\rho^2}}\right]
\end{eqnarray}
It is singular when $S = R$, but a more detailed analysis (see Appendix~\ref{sec:Appendix_B_C2_peak}) shows that this is an artifact of a previous approximation near the focal point $S$ and that in reality the  singular term smears out over one mean free path and takes the finite value $\eta/\ell$ with $\eta = 1.3863$ for $R = S$. 
Since the $C_1$ peak is typically one wavelength in size, the $C_2$ density~(\ref{C2final}) can still be considered  as ``background". 
Nevertheless, as soon as $R \approx S$ within a wavelength an extra factor 2 shows up because of the existence of ``Coherent Backscattering'' of waves released by the pointlike source above and due to the so far neglected second term in Eq.~(\ref{DA}). 
However, because the source is smeared out over a mean free path its peak value is suppressed, hence it will not be discussed in further detail. 

\begin{table}[t]
\begin{tabular}{|c|c|}
  \hline
 depth $z$  &   $B_2(z,0)/B_1(z,0)$    \\
  \hline
  $z=0$ & $3\Omega_A/ 4\pi \times  ( 1+ 2z_0/\ell) ) $ \\
  $z=z_S$ & $ 3\Omega_A/ 4\pi \times  (1+ 4\eta z_S^2/\ell^2)$     \\
  $z= \infty$ &  $3\Omega_A/ 4\pi \times  ( 1 + 2z_S/\ell) $  \\
  \hline
\end{tabular}
\caption{Ratio of $C_2$ and $C_1$ contributions to energy density at point $ R= (z,{\bm \rho}=0)$. The first term ``$1$" in brackets is $B_2^{(3)}$, the second term is $B_2^{(4)}$. Both terms are of same order at $z=0$, but the second term rapidly dominates as the depth $z$ increases. The ratio $B_2/B_1$ is a competition of the limited  angular coverage of the array $\Omega_A/ 2\pi \ll 1$ and the large factor $z_S/\ell$, especially near the focal point $S= (z_s, \bm \rho = 0)$ ($\eta =1.38$).}
\label{table:C1_C2_contributions}
\end{table}
Let us compare the energy density $B_2$ produced by $C_2$ to the density $B_1$ of the $C_1$ background found in Eq.~(\ref{C1background2}), and calculate their ratio as a function of the depth $z$ of point $R$, as is summarized in Table~\ref{table:C1_C2_contributions}. 
{\color{black}The data in the table show that eventually $B_2$ always dominates, as $z_s/\ell$ becomes large enough to compensate for the small factor $3 \Omega_A / (4 \pi)$. }
Near the boundary the ratio $B_2/B_1$ is small and independent of $z_S$.
If the quality of the focusing near $S$ is dominated by $B_2$,  Eq.~(\ref{ratio}) must be replaced by the much larger ratio of background to focal peak that is equal to 
\begin{equation}\label{ratioC2}
     \frac{ B_1 + B_2}{ P_1 (R=S) } \approx \frac{3 \eta}{2}  \frac{1}{(k\ell)^2}
\end{equation}
This is sufficiently small for the focusing to be efficient. 
It is  independent of the depth $z_s$ of the focal point and independent of the angular coverage $\Omega_A$ as well.
Contrary to the $B_1$ energy density discussed in the previous section, Eq.~(\ref{C2final}) contains genuine source of energy at the focal point $S$. The flux emitted by this source is given by $F_S = -D\int d^3\mathbf{r} \bm\nabla^2 B_2^{(4)}(\mathbf{r})$ where the volume integral is to be taken around the point $S$, see Appendix~\ref{sec:Appendix_A_current}. 
We can compare this to the total flux~(\ref{RC1}) leaving the sample and find,
\begin{equation}\label{source2R}
    \frac{F_S}{F_1(z<0)} = \frac{\Omega_A}{\pi}
 \end{equation}
\textit{i.e.}, the relative source strength created by the wave-front shaped signal in the medium depends (only) on the angular coverage of the array and is determined by $\Omega_A$. 
The ratio is small because of the large $C_1$ source created at the incident boundary.

\section{Wave Front Focusing to a point in a Slab}
In the following we show that the above considerations for a half-space change only quantitatively when the focus is performed inside a slab of finite width $L$. The slab geometry is interesting because it allows to investigate the relation between transmission and focus.

In the $C_1$ approximation, the background energy density associated with is still given by Eq.~(\ref{C1background})
but involves the diffuse propagator for a finite slab. Using radiative boundary conditions at both sides $z=-z_0$ and $z= L+z_0$ 
\begin{eqnarray}\label{finiteslab}
&& L(z,z',\mathbf{q})  =\nonumber \\
 &&  \ \frac{12 \pi}{\ell^3} \frac{\sinh q[B-\mathrm{max}(z,z')-z_0]\sinh q[\mathrm{min}(z,z')+z_0]}{q\sinh q B}\nonumber \\
\end{eqnarray}
with $B= L+2z_0$. 
Equation~(\ref{C1background}) becomes,
\begin{eqnarray}\label{backgroundB}
B_1(z,\bm \rho)&= & \frac{9|F|^2N^2 }{(k\ell^2)} \frac{z_0^2}{\Omega_A} \int d^2\mathbf{q}\,  e^{i\bm q \cdot \bm \rho}   \times \nonumber \\
    &&  \frac{\sinh q(B -z_S -z_0)\sinh q(B -z -z_0)}{\sinh^2 q B}
\end{eqnarray}
valid for  $z_S,B  \gg z_0$.
With the expression for $L(z,z',\bm q)$ for the slab,  the peak value  in Eq.~(\ref{peak}) for a half-space is modified by
   \begin{eqnarray}\label{peakslab}
 P_1(S,R)   =  9 |F|^2 N^2  P(S,R)^2  \frac{z_0^2}{\ell^2}\left(1 -\frac{z_S+z_0}{B} \right)^2
 \end{eqnarray}
which in a finite slab  decreases smoothly with  the depth of $S$. The   ratio of background to peak can be written as,
\begin{equation}\label{ratioslab}
    \frac{B_1(R=S)}{P_1(R=S)}= \frac{2\pi/[2k(z_s+z_0]^{2}}{\Omega_A} R\left(\frac{z_s+ z_0}{B}\right)
\end{equation}
The function $R(\tau_S)$ rises from the value $R(0)=1$ for the half-space  to $R(1/2) = 2.772 $ in the middle of the slab to $R(1) = 7.207 $ at the transmitting boundary. 
The same function $R$  modifies the transverse width of the background energy density $\langle\rho^2 \rangle_1 $ around the focus $S$   given by Eq.~(\ref{surfaceC1}) for a half-space:  $\langle\rho^2(z=z_S) \rangle _1\rightarrow \langle\rho^2 \rangle_1/ R\left({z_s+ z_0}/B\right) $. The finite thickness of the slab thus enhances the background and suppresses the transverse size.  
Nevertheless, the ratio $B_1/P_1$ still decays essentially as $1/(kz_S)^2\Omega_A \ll 1$. 


The $C_2$ background density for the finite slab is given by Eq.~(\ref{eq:C2simple}), though with only an additional first term due to the transmitting boundary $z_H=L$, and which cancels for the same reason due to the radiative boundary condition. 
If we neglect all powers in $\ell/B$ we see that the third term $B_2^{(3)}$  only contributes when the Hikami box resides in the skin layer so  that the relation $B_2^{(3)}/B_1 = 3 \Omega_A/4\pi$ for the half space continues to apply for the slab. The last term of Eq.~(\ref{eq:C2simple}) becomes equal to
\begin{eqnarray}\label{C24slab}
 B_2^{(4)}(z, \bm\rho) &&= \frac{9|F|^2 N^2}{(k\ell)^2} \frac{3 \pi z_0^2}{\ell} \rho(z,z_S,\bm\rho)\times \nonumber \\
&& \left[\left( 1- \frac{z+z_0}{B} \right)^2 + \left( 1- \frac{z_S+z_0}{B} \right)^2\right]
\end{eqnarray}
where the $ \rho(z,z_S,\bm\rho)$ is the energy density  emitted by a diffuse point source at position $z=z_S,\bm \rho=0 $ in the slab, given by
\begin{eqnarray}\label{slabsource}
  &&   \rho(z,z_S,\bm\rho)=    \int \frac{d^2 \mathbf{q}}{(2\pi)^2} e^{i\bm q\cdot \bm\rho} \times \nonumber \\
  && \frac{\sinh q[B-\mathrm{max}(z,z_S)-z_0]\sinh q[\mathrm{min}(z,z_S)+z_0]}{q\sinh q B} \nonumber
\end{eqnarray}
This expression diverges at the focal point $S$ because an energy source is created that is in reality smeared out over a mean free path.  The power emitted by this source is slightly lowered for a finite slab and Eq.~(\ref{source2R}) is modified by,
 \begin{equation}\label{source2slab}
    \frac{F_S}{F_1(z<0)} = \frac{\Omega_A}{\pi}\left( 1 - \frac{z_S+z_0}{2B}\right)
 \end{equation}
This ratio slowly decays to $\Omega_A/2\pi$ as the focal point $S$ approaches the transmitting boundary. 

The varying background energy density in the  slab will have a non-zero transmission coefficient. Because of the wavefront-shaped incident beam this transmission will not necessarily decay as $1/B$.
We shall assume the slab to be optically thick enough so that $\ell/L \ll \Omega_A/\pi \ll 1$. 
The transmission induced by $C_1$ is proportional to $(\ell/L)/ \Omega_A$, so that $C_2$ dominates, and in particular the source term $B_2^{(4)}$ given by Eq.~(\ref{C24slab})
\begin{eqnarray*}
  F_2^{(4)}(L) &=&  -D\int d^2\bm\rho\partial_z {B_2^{(4)}(z, \bm\rho) }\\
  &=& 9 \frac{|F|^2N^2}{(k\ell)^2}  {\pi z_0^2v_E} \left( \frac{z_S + z_0}{B}\right) \left( 1 - \frac{z_S}{B}\right)^2
\end{eqnarray*}
where we neglected powers of $z_0/B$. For a finite size of the slab, the reflected current in Eq.~(\ref{RC1}) achieves an extra factor
$(1 - z_S/B)$ due to the presence of $L(0, z_S)$. 
The normalized transmission is thus equal to
\begin{equation}\label{Trans}
    T = \frac{\Omega_A}{2\pi } \frac{ {z}_S + z_0 }{B}\left(1- \frac{z_S+z_0}{B}\right) \leq \frac{\Omega_A}{8\pi }
\end{equation}
By focusing to a point in the slab, the transmission takes a finite value and takes its largest value $T = \Omega_A/8\pi$ independent of thickness $B$ when we focus in the middle of the slab. 
 
On one hand it is surprising to find that WFS to a point facilitates to transmit - via non-Gaussian $C_2$ correlations - in a non-Ohmic way. 
On the other hand, the maximum transmission is far from the optimal transmission $T = 1$ established from random matrix theory in quasi-1D samples. 
This is clearly due to the large $C_1$ reflection coefficient that stems from the traditional source created near the incident boundary. The modes with optimized transmission somehow find a way to suppress this source.  Note also that the energy density expressed by Eq.~(\ref{C24slab}) is not mirror-symmetric, not even when $z_S=B/2$. 
The density associated with optimized transmission must be mirror-symmetric in the plane $z=B/2$ (see argument below). 
This is clearly due to our choice to put the time-reversal array on the left, and not on the right, and can be fixed by using identical, independent arrays on both sides.

\begin{figure*}[htbp]
\includegraphics[width=0.995\textwidth]{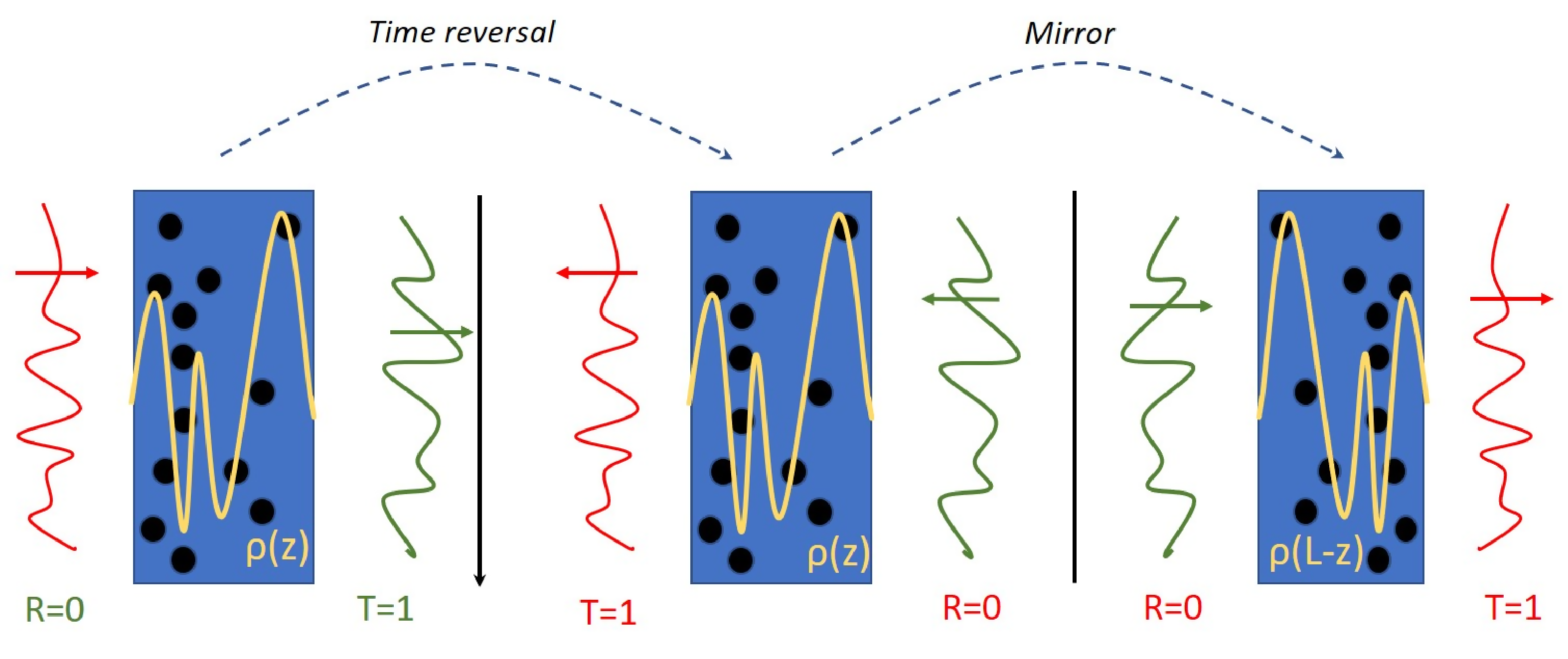}
\caption{wavefront shaping and the energy density inside a slab.
Left: if we optimize the total transmission, it is unity ($T = 1$), whereas total reflection vanishes ($R = 0$) and the energy density $\rho(z)$ varies wildly.
Center: if we apply time-reversal, the incident pattern (from left panel) is incident from the back surface without reflection ($R = 0$) and perfectly transmitted ($T = 1$) to the left entrance surface.
The energy density $\rho(z)$ has the same pattern as it is not affected by time reversal.
Right: if we apply a mirror operation, the incident wavefront from the center image is incident from the left, enters perfectly with ($R = 0$) and transmits perfectly ($T = 1$) to the right.
The energy density $\rho(z)$ is mirrored with respect to left and center since the sample structure has been mirrored.
Averaging over configurations yields a mirror symmetric energy density, hence $\rho_M (z) = \rho_M(L-z)$.
}
\label{fig:cartoon_symmetry}
\end{figure*}
 
 We conclude that the focalisation to a point in the slab is not equivalent to optimizing transmission. Nevertheless, focusing to a point enhances the transmission far beyond the Ohmic expectation $\ell/L$, due to the presence of a source of energy  inside the medium.

\section{Towards Optimized Transmission}\label{sec:optimized_transmission}
In this section we develop the idea that optimizing transmission is related to the creation of an energy source inside the slab. 
We will show that different propositions for the energy density profile lead to quite different profiles for the sources. 
Two general arguments are made.

The first argument concerns the spatial symmetry of the density. 
The procedure is to first optimize the WFS for each different realization to find perfect transmission and next to average the associated energy density inside the slab over all realizations, and with equal incident power. Having found a wavefront that gives full transmission ($T = 1$) for a given realization of the disorder in the slab, see Figure~\ref{fig:cartoon_symmetry}~(Left), the time-reversed operation reproduces the same process, see Figure~\ref{fig:cartoon_symmetry}~(Center), gives again an optimal transmission ($T = 1$) with the same density profile.
More precisely, this follows from the Stokes relation for the complex transmission matrix for the reversed process, $\tilde{t}$ obeys $\tilde{t}_{nm} = t_{mn}$.
If we next perform a mirror operation of both wave and disorder, see Figure~\ref{fig:cartoon_symmetry}~(Right), we have constructed an incident wavefront that has the inverted energy density $\tilde {\rho}(x,y,z) =  {\rho}(x,y,L-z)$ with respect to the initial process, with also maximum transmission.
Assuming that we perform a perfect average over disorder, both energy densities will occur with same statistical weight.
The energy density, averaged over disorder \emph{given} optimal transmission $T=1$,  must be  \emph{symmetric }about the central plane of the slab: $\rho_M(z) = \rho_M(L-z) $. 
Of course, this argument relies on the symmetry of the slab geometry and does no longer apply when the geometry itself breaks mirror symmetry~\cite{Koirala2017PRB}. 

\begin{table*}[htbp]
\begin{tabular}{|c|c|c|c|c|c|}
  \hline
  model &  virtual source  & energy density &  $\rho_M(n)$ & energy & $\rho_M(L/2)/\rho_M(0)$ \\
  \,  &  [$S/B$]  & [$BS/D$] & & [$B^2S/\pi^2 D$] & [$B/\ell$]  \\
   \hline \hline
  delta & $ \delta(\tilde{z}-1/2)$  &  $\frac{1}{4}(1 - |1 -  {2\tilde{z}}|)$ &  $4 (-1)^{n-1} / \pi (2n-1)^2$  & $1.234$ & $0.750$ \\
  log-singular & $ -\frac{\pi}{4 G} \log \tan  \frac{\pi}{2} \left|\tilde{z}-\frac{1}{2} \right|$  &  not ana &  $(-1)^{n-1}  /G(2n-1)^3$ &$1.080$ & $0.548$ \\
  $n=1$ & $\frac{\pi}{2}\sin \pi \tilde{z} $  &  $ \frac{1}{2\pi }\sin \pi \tilde{z} $&  $\delta_{n,1}$ & $1.000$  & $0.477$  \\
best fit & $\frac{1-\alpha (3\tilde{z}^2 -3\tilde{z}+1)}{[1-\alpha \tilde{z}(1-\tilde{z}) ]^3 }$ & $\frac{1}{2} \frac{\tilde{z}(1-\tilde{z})}{1- \alpha \tilde{z}(1-\tilde{z}) }$ &
$0.9966$, $-0.00295$ & $0.996$ &   $0.295$  \\ 
($\alpha = 4 - \pi$) & \, & \, & $+ \mathcal{O}(1)/(2n-1)^{3.36}$ &\, &\\    
  Flat~\cite{} ($\alpha = 0$)  & $1$  &  $\frac{1}{2}{\tilde{z}}\left( 1 - {\tilde{z}}\right)$ &  $8/(\pi^2 (2n-1)^3) $  & $0.822$ & $0.375$ \\ \hline 
  symmetric & $ B\delta(z)/2  $ & $ \tilde{z}/2 \  (z < 0)$ &  $ \sin\pi(2n-1)\tilde{z_0} $ &  $\pi^2/2 $ &  \\
    normal    & $+$ &  $\tilde{z_0}/2  \   (0 < z < L)$ & $\times$  &  $\times $ & $\ell/B  $ \\ 
   incidence &  $B\delta(L-z)/2 $  &  $(1- \tilde{z})/2 $ \   $(z > L)$ & $ 4/\pi (2n-1)^2$   & $\tilde{z_0} -\tilde{z_0}^2$  &  \\ 
 \hline
\end{tabular}
\caption{Some properties of  6 different models for a  3D slab integrated over transverse dimensions $\bm\rho$ or alternatively for a quasi-1D geometry. The  total energy rate of all sources is normalized to $S$.  We have abbreviated $\tilde{z}\equiv (z+z_0)/B$, $B=L+2z_0$, $\tilde{z_0} = 2\ell/3B $,  and  $G= 0.915965..$ is Catalan's constant. The ``best fit" model was used by Ref.~\cite{Koirala2017PRB} with Fourier coefficients that decay relatively  fast and roughly as $1/(2n-1)^{3.36}$. The symmetric plane wave source corresponds to  equal plane waves incident on both sides creating sources near both boundaries.}
\label{table:6_different_models}
\end{table*}

The second argument, detailed in Appendix~\ref{sec:Appendix_A_current}, is that optimization to a point $S$ in the medium produces an energy density whose average obeys a diffusion equation with the usual radiative boundary conditions at both sides, with same diffusion constant, and with some source related to the focal point proportional to the incident power. 
Optimizing to a focal point is not equivalent to optimizing transmission, but let us speculate that the argument also applies for optimized transmission. 
This feature is confirmed by numerical simulations~\cite{genack1}.
The energy density averaged over different optimizations of transmission  $\rho_M(z,\rho)$, must then be a superposition of the complete set of eigenfunctions of the diffusion equation, with a source profile to be determined~\cite{Ojambati2016OE}. 
Let us ignore the complication of transverse energy profile, and integrate either over $\bm\rho$ or restrict to a quasi-1D geometry. 
The energy density would then be ($B=L+2z_0$),
\begin{equation}\label{diff1}
  \rho_M(z) = \frac{B S}{2 \pi D} \sum_{n=1}^\infty \rho_M(n) \sin \frac{ (2n-1)\pi (z+z_0)}{B}
\end{equation}
featuring only the \emph{symmetric }modes with odd $n$, and a front factor that depends on the source power $S$, including the total amount of energy $S$ delivered by a hypothetical source, and with dimensionless coefficients  $\rho_M(n)$ that determine the spatial density profile. 
For a \emph{single} realization of disorder, no rigorous relation exists between density gradient and transmission and optimization leads to {zero reflection $R = 0$ and perfect transmission} $T = 1$.
After averaging however, the diffusion picture emerges and the gradient of $\rho_M$ at the boundaries determines the average outgoing flux, equal on both sides, hence $R = T$ and $R + T = S$. 
The source density $S(z)$ is then given by 
\begin{equation}\label{diff2}
S(z) = {\frac{\pi S  }{2B }}\sum_{n=1}^\infty \rho_M(n) (2n-1)^2 \sin \frac{ (2n-1)\pi (z+z_0)}{B}
\end{equation}
{which is thus necessarily also} symmetric around $z_S = L/2$. 
The extra factor $(2n-1)^2$ implies the resurrection of high-order eigenfunctions in the Fourier expansion for the source that are not all positive-definite. 
Alternating signs with $\rho_M \sim (-1)^{n+1}$ generate more weight in the center since for even $n$ the eigenfunctions are all negative at the center of the sample. 

In Table~\ref{table:6_different_models} we consider 6 different normalized  sources. 
They all share positive energy density \textit{and} positive source density. 
The ``flat" model was previously discussed by Davy \textit{et al.} and corresponds to a homogeneous source density~\cite{Davy2015NatComm}. 
This model was generalized by Koirala \textit{et al.} to the ``best fit" model and the best fit to numerical simulation was obtained for $\alpha\approx 4 - \pi$~\cite{Koirala2017PRB}. 
Furthermore, the simplest model ``$n=1$" keeps only the first eigenfunction, the only symmetric one that is positive definite.  
Finally, the ``symmetric normal incidence  model" is associated with two equal sources close to both boundaries. This model clearly behaves differently from the others, because it  decays slower with $n$ and the Fourier coefficients explicitly depend on $\ell/B$.

\begin{figure}[htbp]
\includegraphics[width=\columnwidth, angle=0]{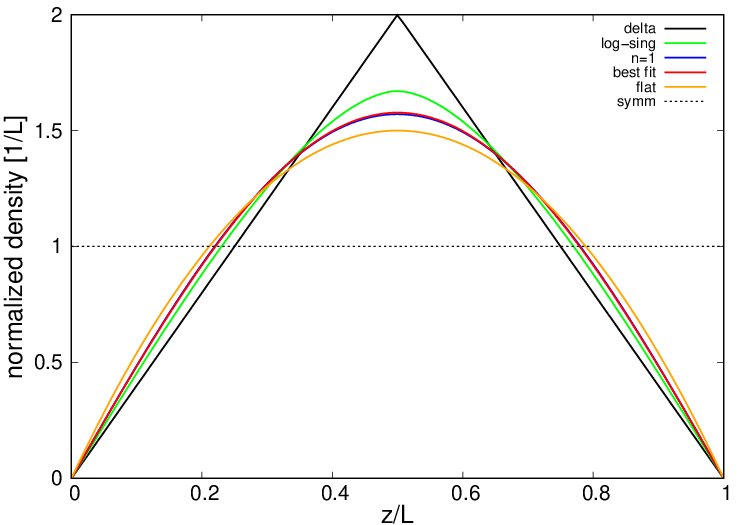}
\includegraphics[width=\columnwidth, angle=0]{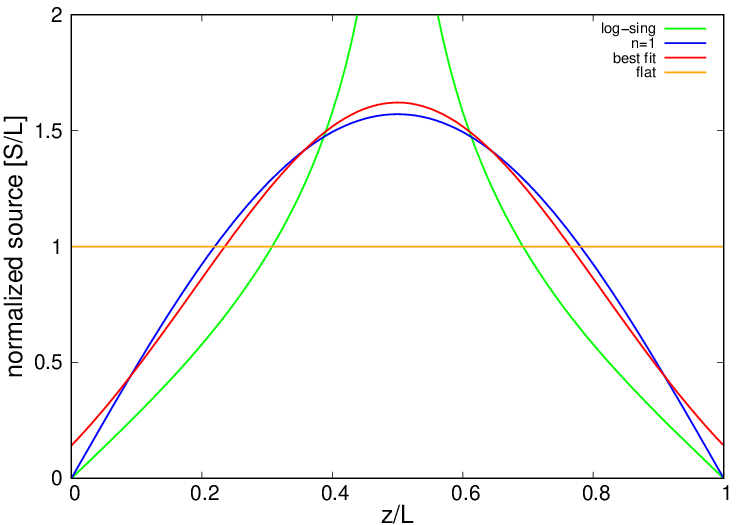}
\caption{Normalized energy densities (top) and normalized sources (bottom) for the six different models that are specified in Table~\ref{table:6_different_models}. }
\label{fig:energy_source_density}
\end{figure}
The assumption of a positive-definite, mirror-symmetric source density filters out many solutions but clearly does not fix the profile. 
The total source power is normalized to $S$ by imposing $\sum_{n=1}^\infty \rho_M(n) (2n-1) = 1$ so that $\rho_M(n)$ must decay at least as fast as $1/(2n-1)^3$. 
This would explain partly that different models for the density profile in Table~\ref{table:6_different_models} have very similar shapes, especially when they are each normalized, see Figure~\ref{fig:energy_source_density}.  
An extreme case, referred to as ``delta" having  a point source in the middle of thee slab can be clearly disqualified by numerical simulations that favor the ``best fit model". 
In Figure~\ref{fig:energy_source_density} we see that the {source distributions} vary strongly from one model to the other. 
The models ``$n=1$'' and ``best fit'' have rather similar sources but hardly distinguishable energy densities. 

This discussion suggests that in order to discriminate between different models for optimized transmission, one should focus on the source density profile, rather than on the energy density. 
The ``$n=1$" model seems to be an accurate candidate, but numerical simulations do reveal the existence of higher modes, $2 \%$ of the energy according to~\cite{Ojambati2016OE}. 
The ``best-fit model" has a finite source density near the boundaries, which is physically reasonable. 
It is also remarkable that the very small value for $\rho_M(n=2)$ is $10 \times$ smaller than the one for the flat model. 
Finally we can investigate the total  energy in the slab associated with optimized transmission. Total energy for optimized transmission can be related to average dwell time for the waves emitted by the source to reach the boundary~\cite{epjb}. For $R=0$, $T=S=1$, 
\begin{eqnarray}\label{eq:friedel2}
  \int d^3 \mathbf{r} \rho_M(\mathbf{r}) =   \frac{d\phi_T}{d\omega}(\bm\rho)  = \frac{B^2}{\pi^2D}  \sum_{n=1}^\infty \frac{\rho_M(n)}{2n-1} 
\end{eqnarray}
The  dwell time itself can be optimized~\cite{Arthur2} or manipulated~\cite{Sarma2016PRL} as well as the closely related delay time~\cite{Rotter2017RMP}. 
We infer from Table II that the stored energy varies only weakly among the different models for optimized transmission. 
This is due to the fact that the Fourier components in Eq.~(\ref{eq:friedel2}) decay as fast as $1/(2n-1)^4$. 
Because phase is measurable after optimizing transmission, this could be an opportunity to measure the diffusion coefficient in optimized transmission.   

\section{Conclusion}\label{sec:conclusion}
In this work we have applied mesoscopic speckle theory to describe the energy density created by a wavefront-shaped incident signal that focuses on a point in the random medium. 
The focus is determined by the short-range $C_1$-speckle whereas the long-range $C_2$ speckle creates a background energy density that dominates deep inside the medium. 
This part also generates an energy source inside the medium. 
The focus is due to constructive interference between incoming and outgoing spherical waves, like in time-reversal experiments, and not to the source. Because $C_1$ speckle creates a source near the incident boundary, much like an incident plane wave usually does in radiative transfer, the focusing to a point does not optimize transmission. 
We have developed the idea that optimizing transmission removes this $C_1$ source and creates a mesoscopic energy source inside the sample that is mirror-symmetric. 
Different models for this source produce quite similar  profiles for the energy density, close to the first eigenmode of the diffusion equation as observed in numerical simulations. 
The red curve labeled ``best fit"  in Figure 4b represents  undoubtedly the energy source that optimizes transmission and is close yet not equal to the lowest diffusion mode. 
A major challenge exists to understand this profile from first principles.

\begin{acknowledgments}
We thank Sergey Skipetrov for useful discussions. 
WLV thanks the CNRS for a fellowship as guest investigator. 
WLV and AL acknowledge support by NWO-TTW Perspectief program P15-36 ``Free-form scattering optics" (FFSO) in collaboration with TU Delft, TU Eindhoven, and industrial users ASML, Demcon, Lumileds, Schott, Signify, and TNO, as well by NWO-TTW Perspectief program P21-20. 
\end{acknowledgments}

\section{Appendix}
\appendix

\section{Current of wave-front shaped waves}\label{sec:Appendix_A_current}

\subsection{Current of the background signal}
The expressions for the background densities in both the $C_1$ and $C_2$ approximation can be generalized to the spatial correlation function
\begin{equation}\label{spatialcorr}
   \langle \Phi( \mathbf{r}) \Phi(\mathbf{r}')^*\rangle = \int d^3\mathbf{r}_4 f(\mathbf{r}_4) G(4,\mathbf{r})G^*(4,\mathbf{r}')
\end{equation}
with $G$ the bulk Dyson Green's function given in Eq.~(\ref{Gbulk}) and $f(\mathbf{r})$ some real-valued function obtained from correlation functions that varies slowly on the scale of the mean free path. For monochromatic scalar waves the cycled-averaged (radiative) density is proportional to $\rho(\mathbf{r})= |\Phi(\mathbf{r})|^2$. For $\mathbf{r}=\mathbf{r}'$ we neglect spatial variation of $f(\mathbf{r})$ within a mean free path and use Eq.~(\ref{eq:peak0}) giving
\begin{equation}\label{density}
    \rho(\mathbf{r})  = \frac{\ell}{4\pi} f(\mathbf{r} )
\end{equation}
The cycle-averaged current-density is $\mathbf{J}= (c_0^2/\omega) \mathrm{Im } \, \Phi^*(\mathbf{r}) \bm\nabla \Phi(\mathbf{r})$. Using
$\mathrm{Im}\, G^*(\mathbf{r}) \bm\nabla G(\mathbf{r})= k \widehat{\mathbf{r}} |G(\mathbf{r})|^2$, we find from Eq.~(\ref{spatialcorr}), putting $\mathbf{x}= \mathbf{r}-\mathbf{r}_4 $

\begin{eqnarray}\label{currentd}
   \mathbf{ J}(\mathbf{r}) &=&  \frac{kc_0^2}{\omega} \int d^3\mathbf{x} f(\mathbf{r}-\mathbf{x}) \widehat{\mathbf{x}} |G(\mathbf{x})|^2\nonumber \\
&\approx & \frac{kc_0^2}{\omega}  \bm   \int d^3\mathbf{x} (-\nabla f(\mathbf{r})) {\mathbf{x}}  \hat{\mathbf{x}}  |G(\mathbf{x})|^2 \nonumber \\
&=& - \frac{1}{3} \frac{c_0^2 }{v_P}\ell \times \nabla  \rho(\mathbf{r})
\end{eqnarray}
with $v_P$ the phase velocity. 
This implies that the diffusion equation applies despite no matter where $f(\mathbf{r})$ stems from, here from two-particle diagrams, and with the same diffusion coefficient as the one found for the one-particle Green's function (the matter energy density inside scatterers should be treated to find the correct velocity). 
Any source or sink of energy in the medium is characterized by a non-zero value for $ \nabla^2\rho({\mathbf r})$.

\subsection{Current of focused signal}

The average field at a point $\mathbf{r}$ near the focal point $S$ follows from Eq.~(\ref{TRM}),
\begin{equation}\label{focusapp}
    \Phi(\mathbf{r}) = NF\int_1 \int_2  e^{-z_1/\ell} L(1, 2) G(S,2) G^*(2,\mathbf{r}) 
\end{equation}
Recall that the factor $F$ has the same unit as the field unit, and that $|F|^2$ has the dimension of field energy density. This expression contains a rapidly varying part on the scale of the wavelength, as well as an exponential decay of the Green's function $G(x)$ on the scale of the mean free path. Because  $L$ varies slower on this scale we can substitute $\mathbf{x}=\mathbf{r}-\mathbf{r}_2$ and $\mathbf{y}=\mathbf{r}-\mathbf{r}_S$, and expand $L(1,2)$ around $\mathbf{r}$ as,
\begin{equation*}
    \int_2 \rightarrow  \int d^3 \mathbf{x}  G(\mathbf{x}-\mathbf{y}) G^*(\mathbf{x}) \left[ L(1,\mathbf{r}) - \mathbf{x} \cdot \bm{\nabla}_\mathbf{r} L(1,\mathbf{r})\right] 
\end{equation*}
The integral over $\mathbf{r}_1 $ generates $\ell L(0,z, \mathbf{q}=0) \equiv \ell \tilde{L}(0,z)$. The  integral over $\mathbf{x}$ can be performed to find,
\begin{eqnarray}
   \Phi(z) \sim {NF\ell^2} \left[ P(y) \tilde{L}(0,z) +i \frac{\ell}{k} P'(y) \bm \nabla_\mathbf{r} \tilde{L}(1,z) \cdot \hat{\mathbf{y}}\right] \nonumber \\
\end{eqnarray}
with $P(y) = -\mathrm{Im}\, G(y)$. The second term in the expansion is usually identified with current but here represents a bipolar contribution to the angular dependence of the focused field around the point $S$. The current density associated with the focused signal is given by $\mathbf{J} = (c_0/k)\mathrm{Im} \Phi^* \bm\nabla_\mathbf{r} \Phi$, involving the \textit{average} field $\Phi$. The released energy at distance $y$ from the focal point follows from the energy flow through a surface  $A=4\pi y^2 $ around the source,
\begin{equation}
    F(y) = \int  d^2 {\mathbf{A}} \cdot \mathbf{J}(S, \mathbf{y} )
\end{equation}
The derivatives $\bm\nabla_{\mathbf y} ({\mathbf y} P ) $  cancel by parity in the angular integral. The derivative $\partial_z $  survives and we obtain, 
\begin{eqnarray}
    F({y}) &= & 
    - \frac{N^2|F|^2c_0\ell^5}
      {3k^4} 
      y^2 P(y)P'(y)
    \left[
      \bm\nabla_{\mathbf{r}} 
      L(1,\mathbf{r}=S) 
    \right]^2 \nonumber \\
\end{eqnarray}
and decays exponentially with the mean free path. Therefore, the focused signal is not associated with a net source and 
$S=\int d^3\mathbf{r}\,  s(\mathbf{r})=0$.

\section{peak value for diffuse $C_2$ source}\label{sec:Appendix_B_C2_peak}
The Hikami source derived in Eq.~(\ref{eq:C2simple}) is proportional to 
\begin{equation*}
    I(\mathbf{\mathbf{r}},\mathbf{r}_S) = \int d^3\mathbf{r}_2 \int d^3\mathbf{r}_4 |G(\mathbf{r}-\mathbf{r}_2)|^2 \frac{1}{ |\mathbf{r}_2-\mathbf{r}_4|} |G(\mathbf{r}_S-\mathbf{r}_4)|^2
\end{equation*}
with $|G(\mathbf{x})|^2 = \exp(-x/\ell)/(4\pi x)^2$. 
For $\mathbf{r}_S$ and $\mathbf{r}$ separated by more than a mean free path {$(|\mathbf{r}_S - \mathbf{r}| > \ell)$} the two integrals decouple, and using Eq.~(\ref{eq:peak0}), the Hikami source takes the form of a point source: 
\begin{equation}\label{far}
    I(\mathbf{r},\mathbf{r}_S) = \left(\frac{\ell}{4\pi} \right)^2 \frac{1}{ |\mathbf{r}_S-\mathbf{r}|}
\end{equation}
When $\mathbf{r}_S$ and $\mathbf{r}$ are closer this is no longer valid. 
{In the limit that both positions coincide $(\mathbf{r}_S = \mathbf{r})$} we get
\begin{eqnarray}\label{close}
  &&  I(\mathbf{r}_S,\mathbf{r}_S) = \frac{\ell}{(4\pi)^4} \int d^3\mathbf{x} \int d^3\mathbf{y} \frac{e^{-(x+y)}}{x^2y^2}\frac{1}{ |\mathbf{x}-\mathbf{y}|}\nonumber\\
 && \    = \frac{\ell}{(4\pi)^2} 2 \int_0^{\pi/4} \frac{d\phi}{\cos\phi (\cos \phi + \sin\phi)}\nonumber \\
&=& \left( \frac{\ell}{4\pi} \right)^2   \frac{1.3863}{\ell} \nonumber
\end{eqnarray}
This outcome comes down to replacing $1/r$ in Eq.~(\ref{far}) for $r=0$ by $\eta/\ell$, with $\eta = 1.3863$. 

The power density of the source is equal to $\bm\nabla \cdot\mathbf{ J}$. The total power is proportional to
\begin{eqnarray}
  \int d^3\mathbf{r} (-\nabla^2) I(\mathbf{\mathbf{r}},\mathbf{r}_S) = 4\pi \left( \frac{\ell}{4\pi}\right)^2
\end{eqnarray}
This means that the point source in Eq.~(\ref{far}) is in reality smeared out over one mean free path $\ell$ when $\mathbf{r}_S$ and $\mathbf{r}$ are close, without affecting its total power. 

\section{$C_0$ correlations}\label{sec:Appendix_C_C0_correlations}
The $C_0$ correlation was first introduced by Shapiro as a fluctuation of the source power by a nearby scatterer, and yields spatial correlations in intensity of infinite range~\cite{Shapiro1999PRL}. 
It was later shown that $C_0$ correlations are fluctuations of the local density of states (LDOS), to which any source is sensitive~\cite{vanTiggelen2006PRE, Birowosuto2010PRL}. 
We here consider the importance of $C_0$ speckle for the quality of the focusing.
We emphasize that even if the source $S$ is virtual, the focusing at $S$ is affected by nearby scatterers. 
The detection point $R$ is assumed to contain no real detector but fluctuations in LDOS at $R$ do exist. 

\begin{figure*}[htbp]
\includegraphics[width=0.9\textwidth, angle=0]{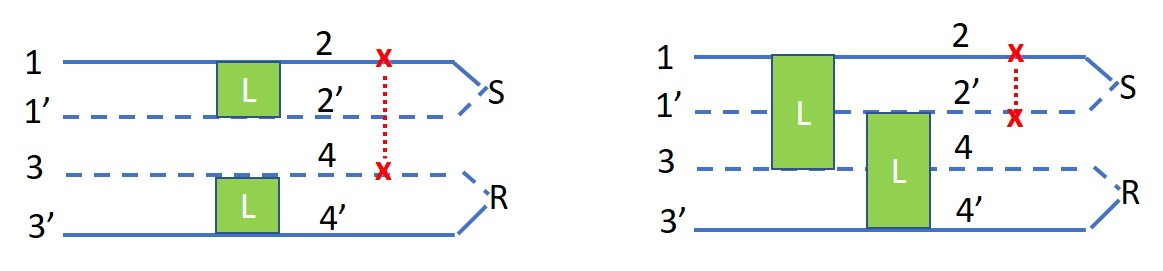}
\caption{$C_0$ contributions to the energy density at position $R$ caused by a scatterer (in red) close to the focus locus $S$. 
Dashed lines denote complex conjugate wave fields. Both actually generate a contribution to the peak, but only the left diagram suffers from decorrelation between the $N$ channels. Complex conjugates of these diagrams exist but are not shown.}
\label{figm3}
\end{figure*}
The two $C_0$ diagrams contributing to this speckle are shown in Figure~\ref{figm3}. 
They describe the perturbation of both background and peak by a scatterer close to $R$ and $S$. 
This implies immediately that both diagrams survive only when $R$ and $S$ are separated by at most one mean free path. 
Both diagrams thus contribute to the focusing and not to the background, and the infinite correlation of $C_0$ does not pertain to $R$.
Let us first consider the righthand side figure. It correlates all incident channels as was the case for the $C_1$ peak in Eq.~[\ref{peak}). The calculation goes as before leading to Eq.(\ref{peak}) with a minor modification of Eq.(\ref{eq:peak0}) that integrates out the positions $2$ and $2'$ near $S$. 
For scatterers "$p$" and scattering matrix $t$ this equation is replaced by,
\begin{eqnarray}\label{C01}
  \left( \frac{\ell}{4\pi}\right)^2 P(S,R)^2  &\rightarrow& 2\times \int_{2, 2'} \sum_p    G(2,p) t G(p, S) G^*(2,R) \nonumber \\
  \,   &\times & G(2',R) G^*(2',p) t^* G^*(p,S)     \nonumber  \\
  \,  =  \left( \frac{\ell}{4\pi}\right)^2 &\times& \frac{8\pi  }{\ell } \int d^3\mathbf{r}_p  P(p,R)^2 | G(p,S)|^2
\end{eqnarray}
where we have replaced $\sum_{p} = \rho \int d^3 \mathbf{r}_p $ and used $\rho |t|^2 = 4 \pi/\ell$. The first factor in this equation is just the one  found for the $C_1$ focusing peak, the second factor stands for the relative $C_0$ correction. The integral over $\mathbf{r}_p$  averages out the oscillation  of the sinc-function $P(p,R)$ and this factor decays as $ \exp(-|R-S|/\ell)/k^2 |R-S|^2$. For $R=S$ we find the $C_0$ factor equal to $\pi/k\ell$ consistent with previous work~\cite{vanTiggelen2006PRE}.

The diagram on the left of Figure~\ref{figm3} can dealt be with similarly, but suffers from decorrelation between the $N$ channels, like the $C_1$ background in Eq.~(\ref{C1background}). 
This time the nearby scatterers $p$ impose a factor different from the one in Eq.~(\ref{C01}), namely 
\begin{eqnarray}
 \mathrm{Re}\, \frac{8\pi  }{\ell } \int d^3\mathbf{r}_p  P(p,R)G^*(p,R) P(p,S) G(p,S) 
\end{eqnarray}
A precise analysis shows that this expression decays exponentially with the distance between $R$ and $S$ as well and thus contributes to the peak. 
For $S=R$ it takes again the value $\pi/k\ell$. 
However, this $C_0$ diagram is  an extra  factor $1/ \Omega_A (kz_s)^2$ smaller than the one expressed by Eq.~(\ref{C01}) because this time the $N$ channels are not all correlated. It can thus be neglected.

We conclude that the $C_0$ correlation contributes to the focusing peak around the virtual source $S$, with a relative weight $\pi/k\ell$, but does not significantly change the picture put forward by the $C_1$ approximation.  Whereas the latter typically predicts a diffraction-limited function $(\sin kx/kx)^2$, the $C_0$ peak decays as $\exp(-x/\ell)/(kx)^2$ around $S$ with an amplitude that is a factor $\pi/k\ell$ smaller.


\end{document}